\documentclass[aps,prl,twocolumn,superscriptaddress,longbibliography]{revtex4-2}
\usepackage[dvipsnames]{xcolor}
\usepackage{graphicx}
\usepackage{amsmath}
\usepackage{amssymb}
\usepackage{newtxtext,newtxmath}
\usepackage{bm}
\usepackage{float}
\usepackage[colorlinks=true,linkcolor=Blue,citecolor=Blue,urlcolor=Blue]{hyperref}
\graphicspath{{figures/}}

\newcommand{\SM}{Supplemental Material}
\begin{document}

\title{Magnetic-texture winding controls fermion-parity switches in an
interacting $p$-wave magnet ring}

\author{Hyun-Yong \surname{Lee}}
\email{hyunyong@korea.ac.kr}
\affiliation{Division of Semiconductor Physics, Korea University, Sejong 30019, Korea}
\affiliation{Department of Applied Physics, Graduate School, Korea University, Sejong 30019, Korea}
\affiliation{Center for Extreme Quantum Matter and Functionality, Sungkyunkwan University, Suwon 16419, South Korea}

\date{\today}

\begin{abstract}
Coplanar magnetic spirals map locally onto uniform spin--orbit-coupled wires,
but on a ring the rotating spin frame can change the electronic boundary
condition. We prove, level by level in every fixed-particle-number sector,
that even and odd texture windings impose boundary phases separated by half
the single-electron flux period. Changing the winding by one also changes
the physical pitch in inverse proportion to the circumference; a symmetric
comparison removes this leading pitch correction. In the number-conserving
topological phase, a global parity constraint then reverses the
fermion-parity assignment of neighboring phase-winding branches.
Density-matrix renormalization-group simulations show this reversal, support
an Ising transition coexisting with a gapless charge mode, and relate the
finite-size parity splitting to an independently calculated charge
stiffness. Magnetic-texture winding thus changes the many-body spectrum
through a global boundary condition that is not determined by the local band
dispersion, providing a closed-geometry, number-conserving probe of
topological pairing.
\end{abstract}

\maketitle

\emph{Introduction.}---%
Models of one-dimensional topological superconductivity established the
connection between pairing and Majorana boundary modes
~\cite{Kitaev2001,AliceaReview2012}. Semiconductor-wire implementations
combine spin--orbit coupling, Zeeman splitting, and induced pairing
~\cite{LutchynSauDasSarma2010,OregRefaelVonOppen2010,Mourik2012}.
In compensated magnets, momentum-dependent spin splittings are even in
momentum in
altermagnets~\cite{Smejkal2022} and odd in momentum in $p$-wave
magnets~\cite{Hellenes2023,Brekke2024}. A metallic spin helix has recently
been reported as a $p$-wave magnet~\cite{Yamada2025}, motivating proposals
for topological superconductivity with Majorana modes and no applied Zeeman
field~\cite{KSP2026,LuoLaw2026,Liu2026,YiAltermagnet2026}. The corresponding
mapping is local: a site-dependent spin rotation
maps a coplanar spiral, bond by bond, onto a uniform wire with odd-in-momentum
spin splitting and a transverse exchange field
~\cite{Braunecker2010,BrauneckerSimon2013,KlinovajaStanoLoss2013,
Kjaergaard2012,MartinMorpurgo2012}. This mapping fixes the bulk dispersion
but not the electronic boundary condition on a ring. After one circuit, the
rotation may return with a minus sign, while adjacent winding numbers also
produce a small change in the local pitch. This distinction is rooted in the
spinor nature of electrons: the classical moments return to their original
orientation after one full turn, while the rotation acting on an electron
changes sign. An open chain imposes no periodicity constraint on the rotated
fermions, so this sign has no separate consequence. On a ring it changes
their boundary condition after one circuit and can reorganize the many-body
spectrum without changing any local coupling. The closed geometry is
therefore essential to the effect. The question is whether an interacting
spectrum can distinguish this discrete boundary sign from the smooth pitch
dependence.
The discrete boundary sign is invisible in the bulk band dispersion but can
appear in observables that depend on taking an electron around the ring. Flux
insertion is particularly useful because it changes the electronic boundary
phase without changing the local magnetic texture.

Particle-number conservation makes that question nontrivial. In a
short-range one-dimensional system, pair correlations decay algebraically
and coexist with a gapless charge mode rather than true superconducting
long-range order. Such a paired system can nevertheless have a gapped neutral
sector with the Ising structure associated with a topological superconductor
~\cite{FLNF2011,Sau2011,ChengTu2011,Ortiz2014,RBA2015,KeselmanBerg2015}.
Interactions can shift the transition and modify boundary properties
~\cite{StoudenmireAlicea2011,TMM2025}, so the response must be formulated in
fixed-particle-number sectors rather than inferred from a mean-field band
invariant. Magnetic textures are known to produce geometric phases and
persistent currents in mesoscopic rings~\cite{LGB1990}, number parity is known
to control the flux response of an interacting Luttinger-liquid
ring~\cite{Loss1992,SinghRoy2020}, and a boundary twist is known to change the
ground-state fermion parity of a number-conserving topological
superconductor~\cite{Ortiz2014,LapaLevin2020,LiuColeSau2019}.

\begin{figure}[t]
\centering
\includegraphics[width=\columnwidth]{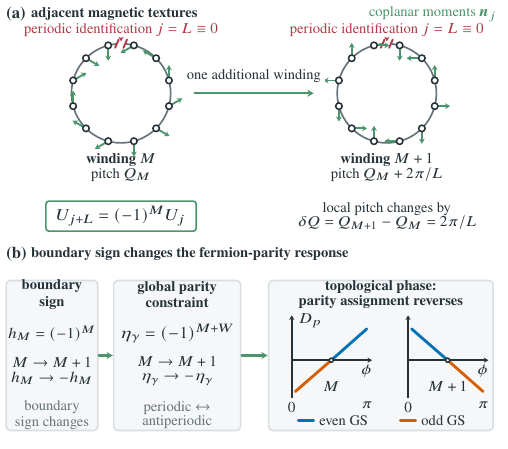}
\caption{\label{fig:model}
\textbf{Boundary condition and fermion-parity response.}
(a) Adjacent coplanar textures have pitches $Q_M$ and
$Q_{M+1}$. The spin rotations that align the textures obey
$U_{j+L}=(-1)^M U_j$; adding one winding reverses this boundary
sign. (b) $W$ denotes the winding of the pair phase, and electron
single-valuedness fixes the neutral-fermion boundary condition
$\eta_\gamma=(-1)^{M+W}$. In the topological phase, $M\to M+1$ reverses the
parity assignment of neighboring phase-winding branches and hence the leading
flux slope of the signed splitting
$D_p=E_{\rm odd}^{\rm corr}-E_{\rm even}^{\rm corr}$. Adjacent physical
textures also differ by a finite-size pitch shift.}
\end{figure}

Here we derive an exact fixed-number relation between magnetic-texture winding and the
electronic boundary phase. At fixed particle number, every level for a given
texture maps exactly onto the corresponding level of the rotated-frame
Hamiltonian. The texture contributes to the boundary phase according to
whether its winding is even or odd, and the two cases differ by half the
single-electron flux period. In the topological phase, the neutral spectrum
and a global
parity constraint convert this boundary-phase shift into a reversal of the
parity assignment between neighboring phase-winding branches. We examine these
distinct statements with number-conserving simulations of
charging-corrected addition energies. Adjacent windings show the parity
reversal, a symmetric comparison cancels the leading pitch correction,
finite-entanglement scaling supports an Ising transition with a gapless
charge mode, and the flux curvature provides an independent estimate of the
finite-size splitting. Figure~\ref{fig:model} summarizes how an additional
texture winding changes the boundary sign and, in the topological phase, the
parity-resolved flux response.

\emph{Interacting $p$-wave magnet ring.}---%
We study the $L$-site ring in Fig.~\ref{fig:model}(a) at density
$n=N/L=1/2$.
\begin{align}
H_M(\phi)&=-t\sum_{j,s}\big(e^{i\phi/L}c^\dagger_{j,s}c_{j+1,s}+\mathrm{h.c.}\big)
-U\sum_j n_{j\uparrow}n_{j\downarrow}\nonumber\\
&+J\sum_j c_j^\dagger\big[\sin(Q_Mx_j)\,\sigma^x
+\cos(Q_Mx_j)\,\sigma^z\big]c_j,
\label{eq:model}
\end{align}
Here $\phi=2\pi\Phi/(h/e)$ is the dimensionless Aharonov--Bohm phase,
$c_j=(c_{j\uparrow},c_{j\downarrow})^{T}$, and
$c_{j+L,s}=c_{j,s}$. We set $t=1$ and use attraction $U=3$ and exchange
strength $J$. The static coplanar spiral has $Q_M=2\pi M/L$, with $x_j=j$ in
lattice units, so the texture winds $M$ times around the ring and the bond
phases $\phi/L$ sum to the total Aharonov--Bohm phase $\phi$. At $n=1/2$,
corresponding to quarter filling of the two spin bands, our reference texture
is the period-four spiral $Q_{M_0}=\pi/2$, or $M_0=L/4$.
Equation~\eqref{eq:model} conserves total charge but no uniform spin component
and contains no external Zeeman field, intrinsic spin--orbit coupling, or
explicit pairing field. The magnetic texture is treated as a static
classical background.

The local rotation $U_j=e^{-iQ_Mx_j\sigma^y/2}$ aligns the exchange field.
For $d_j=U_j^\dagger c_j$, the Hubbard term is unchanged and the
single-particle Hamiltonian is translationally invariant [\SM{}~S1]:
\begin{equation}
\mathcal H(k)=-2t\cos\tfrac{Q_M}2\cos k\,\openone
+\underbrace{2t\sin\tfrac{Q_M}2}_{\textstyle\alpha}\,\sin k\,\sigma^{y}
+J\,\sigma^{z}.
\label{eq:band}
\end{equation}
Equation~\eqref{eq:band} contains an odd-in-$k$ spin splitting
$\alpha\sin k$, with $\alpha=2t\sin(Q_M/2)$, transverse to the uniform
exchange field $J\sigma^z$~\cite{Kjaergaard2012,MartinMorpurgo2012}. At
$J=0$ the splitting can be removed by undoing the local rotation; for
$J\ne0$ it does not commute with the exchange field [\SM{}~S1]. Removing one
term by a spin rotation therefore restores the spatial dependence of the
other. At the reference pitch and density, only the lower hybridized band
crosses the Fermi level, at rotated-frame momenta $k=\pm\pi/2$. Projecting
the attraction onto this band yields an odd-parity intraband pairing channel.
The corresponding paired Bogoliubov--de Gennes Hamiltonian is generically in
class D
~\cite{Kitaev2001,AliceaReview2012}, whereas the microscopic Hamiltonian
remains number conserving. Equation~\eqref{eq:band} fixes the local band
physics but not the boundary condition of the rotated fermions on the ring.

\emph{Texture winding and the boundary condition.}---%
On an open chain the rotation can be chosen site by site, so the spiral is
unwound without a periodicity constraint. On a ring the same rotation maps
every local term, but the rotated fermions must also satisfy a boundary
condition after one circuit. A spin-$1/2$ rotation through $2\pi M$ returns
with the sign
\begin{equation}
U_{j+L}=(-1)^{M}U_j,\qquad h_M=(-1)^{M}.
\label{eq:hM}
\end{equation}
Thus the rotated fermions obey $d_{j+L}=h_Md_j$ and are periodic or
antiperiodic according to the parity of $M$. This discrete sign differs from
the continuously tunable spin rotation accumulated in a generic Rashba ring
~\cite{AharonovCasher1984}.
The combined spin rotation and $U(1)$ gauge transformation
$V_{M,j}=e^{i\pi Mx_j/L}U_j$, defined by $c_j=V_{M,j}f_j$, is single-valued,
$V_{M,j+L}=V_{M,j}$, because its two factors acquire the same sign. It
converts the boundary sign into a $U(1)$ boundary phase and yields the
fixed-$N$ identity
\begin{equation}
\boxed{\;\operatorname{Spec}_N H_M(\phi)
=\operatorname{Spec}_N H_{\rm uni}\big(Q_M,\;\phi+\pi M\big)\;}
\label{eq:spectralidentity}
\end{equation}
(mod $2\pi$), where $\operatorname{Spec}_N$ denotes the fixed-$N$ spectrum
and $H_{\rm uni}$ the uniform rotated-frame Hamiltonian (\SM{}~S1). A
boundary phase of $\pi$ per electron is one superconducting flux quantum; for
a spin-gapped, charge-gapless system this periodicity is exact in the
thermodynamic limit without any pairing field~\cite{SeidelLee2005}.
This identity holds level by level in every fixed-$N$ sector, without a
mean-field or long-wavelength approximation. Both
transformations leave the local interaction unchanged. The discrete boundary
sign therefore produces the exact phase shift $\pi M$ independently of
interaction strength and of the many-body phase.

The exact boundary phase must be distinguished from a comparison of adjacent
physical textures. We extract a signed parity splitting
$D_p^{(M)}(\phi)$ from the addition spectrum by subtracting its smooth
charging background; $D_p>0$ means that the even sector lies lower.
Changing $M$ by one shifts the boundary phase by $\pi$ but also changes the
pitch by $2\pi/L$. Provided both pitches remain in the same topological phase
and no additional crossing between the parity sectors occurs, the scaled
splittings agree after the $\pi$ shift up to a correction of order $1/L$.
Averaging the $M_0-1$ and $M_0+1$ responses cancels the leading correction
about $Q_{M_0}$ (\SM{}~S1).
Therefore the boundary-phase contribution is exact, while the comparison of
adjacent physical textures approaches this relation with increasing $L$.

The boundary sign depends on the total winding rather than on a uniform
pitch. A nonuniform coplanar texture with the same total winding gives the
same periodic or antiperiodic boundary condition, although the rotated
couplings and the local energy response need not be uniform. Thus the exact
boundary-condition statement extends beyond the ideal spiral used in the
simulations (\SM{}~S1).

\emph{Low-energy theory and global parity constraint.}---%
At small $J$, Eq.~\eqref{eq:model} connects continuously to a
Luther--Emery liquid with a gapless charge mode and a gapped, topologically
trivial neutral sector~\cite{LutherEmery1974,Giamarchi2004}. For spinless
electrons in a single channel no separate weakly paired phase survives in an
isolated wire~\cite{KaneSternHalperin2017}; here the two spin components
supply the neutral sector whose gap distinguishes the phases. The transition
is represented by a neutral Majorana mass $m(J)$ changing sign; other
massive neutral modes have been integrated out. This low-energy description
follows the number-conserving treatments of
Refs.~\cite{FLNF2011,RBA2015,KeselmanBerg2015}; its motivation and
limitations at $U=3$ are discussed in \SM{}~S2. Near the decoupled fixed
point, the leading Hamiltonian is
$H_{\rm low}=H_\rho+H_{\rm Ising}$,
\begin{align}
H_\rho&=\frac{u_\rho}{2\pi}\int_0^L\!dx\,
 \Bigl[K_\rho(\partial_x\theta_\rho)^2
+K_\rho^{-1}(\partial_x\phi_\rho)^2\Bigr],\nonumber\\
H_{\rm Ising}&=\int_0^L\!dx\,\Bigl[ 
\frac{i v_I}{2} \sum_{s=L,R}
\eta_s \chi_s\partial_x\chi_s
+i m(J)\chi_L\chi_R\Bigr].
\label{eq:low}
\end{align}
Here $\eta_L=+1$ and $\eta_R=-1$, while $u_\rho$ and $K_\rho$ are the charge
velocity and Luttinger parameter, and $\chi_{R,L}$ are neutral Majorana
fields with velocity $v_I$. The physical pair phase is
$\Theta=\sqrt{2}\theta_\rho$. Within the neutral Ising theory, the sign of
$m$ distinguishes the trivial and topological phases, and $|m|$ sets the
neutral-gap scale. Symmetry allows a marginal coupling between the charge and
Ising sectors, which can alter the transition in some parameter regimes
~\cite{Alberton2017}; its explicit form is given in \SM{}~S2. The two sectors
therefore need not remain exactly decoupled in the microscopic model, and the
numerical scaling below probes the proposed continuous transition. For
$m\ne0$ the neutral sector is gapped and the charge mode gives $c=1$. At a
continuous Ising transition, a critical Majorana mode adds $c=1/2$ and gives
$c=3/2$~\cite{RuhmanAltman2017}. The central charge supports the proposed
critical degrees of freedom but does not identify which gapped phase is
topological; that distinction comes from the winding-dependent parity
response.

Although Eq.~\eqref{eq:low} separates the two sectors in the bulk, the
physical states on a ring obey a global parity constraint. Write the
low-energy electron in the occupied band as
$\psi_-\sim e^{i\Theta/2}\gamma$, where the neutral fermion $\gamma$ has
chiral components $\chi_R,\chi_L$. Define its boundary condition by
$\gamma(x+L)=\eta_\gamma\gamma(x)$. For an integer winding $W$ of the pair
phase,
$\Theta(x+L)=\Theta(x)+2\pi W$, the charge factor changes by $(-1)^W$, and
the texture contributes $(-1)^M$. Single-valuedness of the electron
therefore requires $\eta_\gamma=(-1)^{M+W}$. Within the paired low-energy
sector, charge excitations change $N$ by two, so the neutral sector carries
the electron-number parity. The neutral-fermion boundary condition and its
fermion parity are nevertheless distinct: either boundary condition admits
both parities, and the neutral spectrum determines which state is lowest. In
the topological phase, the two boundary conditions have lowest states with
opposite parities. Denoting the fermion parity $(-1)^N$ of the lowest state
at fixed $(M,W)$ by $P_{\rm gs}(M,W)$ gives
\begin{equation}
P_{\rm gs}(M,W)=\kappa\,(-1)^{M+W}.
\label{eq:parityconstraint}
\end{equation}
The factor $\kappa=\pm1$ is a nonuniversal microscopic sign. It is
fixed within a connected topological phase as long as no additional crossing
between the parity sectors occurs. Changing either $M$ or $W$ by one then
reverses the relative parity assignment
[Fig.~\ref{fig:model}(b); \SM{}~S3].

\begin{figure}[t]
\centering
\includegraphics[width=\columnwidth]{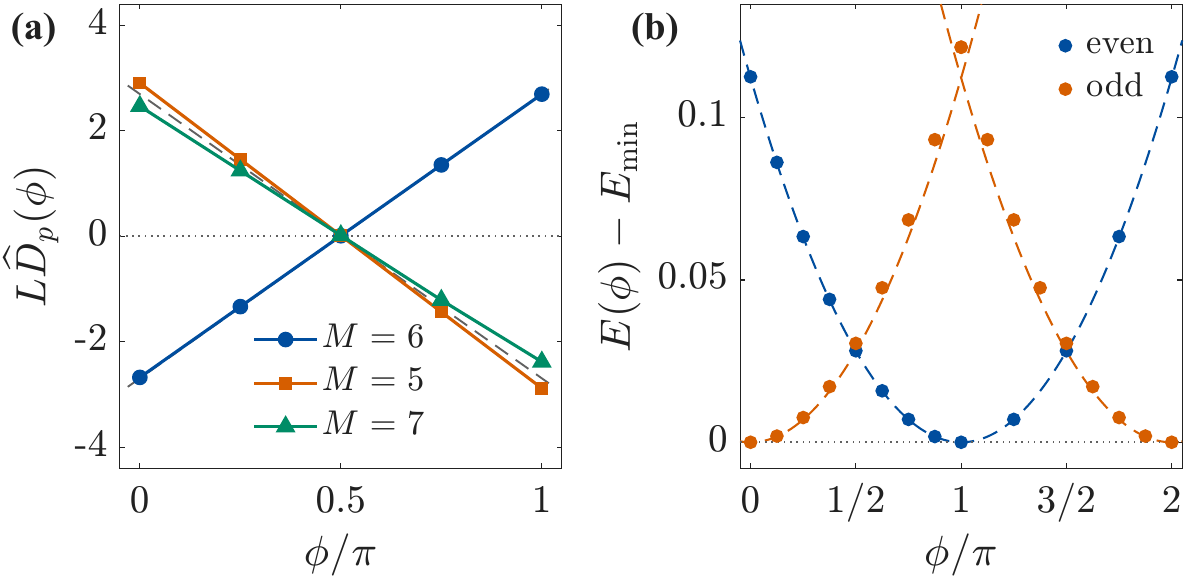}
\caption{\label{fig:response}
\textbf{Reversal of the parity assignment between adjacent texture windings.}
(a) The scaled, charging-corrected five-point estimator
$L\widehat D_{p,M}$ at $L=24$ for adjacent texture windings $M=5,6,7$:
adding one winding reverses the parity assignment of the phase-winding energy
branches.
(b) Parity-resolved energies at the reference winding
$M_0=6$ ($L=24$): the even- and odd-parity minima are displaced by $\pi$ in
$\phi$ and approach the same bulk stiffness, although their finite-size
curvatures differ (dashed parabolas). Numerical data and convergence analysis
are given in the \SM.}
\end{figure}

The $\pi$ shift is present on both sides of the transition; what differs
is whether it has an observable consequence. In the
topologically trivial phase, the lowest opposite-parity state is separated
by a finite neutral gap, so $D_p^{(M)}(\phi)$ is dominated by that gap and
varies weakly with flux. Shifting a nearly flat response by $\pi$ leaves its
sign unchanged, and no interchange of parity assignments occurs. Changing $W$
cannot remove this neutral excitation. In the topological phase, the lowest
state of opposite parity instead lies in an adjacent $W$ sector. For rings longer than
the neutral correlation length, this state has lower energy once its
$O(1/L)$ phase-winding energy falls below the neutral gap. The boundary sign
then determines the relative parity assignment. This explains why the exact
boundary phase is not by itself a topological diagnostic: it is present on
both sides of the transition, but only in the topological phase does the
neutral spectrum allow an opposite-parity state whose cost decreases as the
ring grows. On
$0\leq\phi\leq\pi$, the leading Gaussian charge theory gives one parabola per
phase-winding sector,
$E_{\rm wind}^{(M)}(W;\phi)\simeq
[\rho_s(Q_M)/2L](\phi-\pi W)^2$, where $\rho_s(Q_M)$ is the leading bulk
charge stiffness, or the energy cost of a slow phase twist.
Equation~\eqref{eq:parityconstraint} assigns opposite parities to neighboring
$W$, so
their leading difference is linear in flux,
\begin{equation}
D_p^{(M)}(\phi)\simeq s_M\,\frac{\pi\rho_s(Q_M)}{L}
       \Bigl(\phi-\frac{\pi}{2}\Bigr),\qquad
s_{M+1}=-s_M .
\label{eq:parityflow}
\end{equation}
Here $s_M=\pm1$ depends on the overall microscopic parity assignment and
alternates with $M$.
At finite $L$, the even- and odd-sector curvatures $\rho_e(L)$ and
$\rho_o(L)$ can differ because of subleading corrections, but both approach
$\rho_s$. For the parity-resolved branches $E_{e/o}(\phi,L)$, we define
$\rho_{e/o}(L)=L\,\partial_\phi^2E_{e/o}|_{\phi=\phi_{e/o}^{\min}}$.
These curvatures provide an independent determination of the phase-winding
energy. At zero flux and even $M_0$,
Eq.~\eqref{eq:parityflow} gives
$|D_p^{(M_0)}(0)|L\simeq\pi^2\rho_e(L)/2$; definitions and finite-size
corrections are given in \SM{}~S3.

\emph{Winding-dependent parity response.}---%
We evaluate the signed parity splitting with number-conserving density-matrix
renormalization group (DMRG) calculations
~\cite{White1992,Schollwoeck2011,TeNPy}. The five-point estimator
$\widehat D_{p,M}$ removes the leading smooth charging contribution; its
definition and convergence are given in \SM{}~S3, and the residual smooth
background it leaves is derived in \SM{}~S6. It is evaluated from adjacent fixed-particle-number sectors around the
reference density, so no pairing field or grand-canonical construction is
introduced. In the topological phase at fixed $M$, flux insertion produces
the usual switch of the ground state between neighboring phase-winding
branches of opposite parity. The texture-dependent result is that the slopes
for $M=5,6,7$ alternate
[Fig.~\ref{fig:response}(a)]. At $J=1.5$ the three flux dependences are nearly
linear and cross close to $\phi=\pi/2$. Number conservation prevents two
opposite-parity branches from hybridizing when they meet, but it neither
requires a crossing nor fixes its position. This behavior is analogous to the
fixed-parity spectral flow underlying the fractional Josephson effect
~\cite{KwonYakovenko2004}, although the present observable is a
number-conserving addition spectrum. The slope alternation persists for
$L=16,20,24,32$ and is independent of the microscopic sign $\kappa$.
Averaging the $M=5$ and $M=7$ responses cancels the leading smooth pitch
correction about $M_0=6$; the residual finite-size dependence is given in
\SM{}~S5. This symmetric comparison is important because it isolates the
alternating boundary-condition contribution from the smooth dependence of
the local band parameters on pitch.

The parity-resolved energies form displaced parabolas
[Fig.~\ref{fig:response}(b)]. Although their finite-size curvatures differ,
both approach the same bulk charge stiffness. The stiffness obtained from
the even-sector flux curvature predicts the large-$L$ value of
$|\widehat D_{p,M}(0)|L$ within about $1\%$ (\SM{}~S5). This agreement
supports the identification of the finite-size splitting with the
phase-winding energy. The curvature is obtained within a single parity sector
and does not use the even--odd splitting, so it supplies an independent check
of the relevant energy scale; by itself, it is not a topological diagnostic.

\begin{figure}[t]
\centering
\includegraphics[width=\columnwidth]{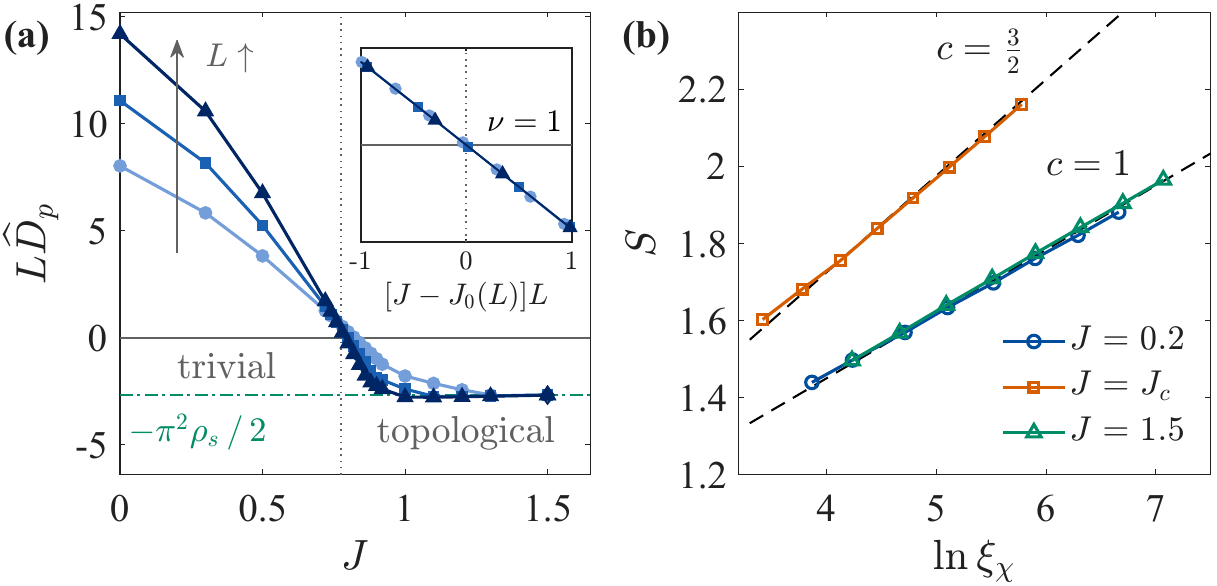}
\caption{\label{fig:validation}
\textbf{Parity scaling and the Ising transition.}
(a) Scaled zero-flux parity-splitting estimator $L\widehat D_{p,M}$ at the
reference pitch $Q=\pi/2$ for $L=16,24,32$, with additional $J=1.5$ points
for $L=40,48$: a finite neutral gap makes it grow with $L$ in the trivial
phase, whereas an $O(1/L)$ phase-winding energy makes it saturate near
$-\pi^2\rho_s/2$ in the topological phase. Inset:
centering each size at its size-dependent zero $J_0(L)$ gives a data collapse
consistent with Ising finite-size scaling with $\nu=1$. (b) Entanglement
entropy $S$ versus finite-entanglement correlation length $\xi_\chi$; the
fitted slopes are consistent with
$c\simeq1,3/2,1$ across the transition. Details are in the \SM.}
\end{figure}

\emph{Ising transition.}---%
The estimator $\widehat D_{p,M_0}(0)$ distinguishes two finite-size energy
scales [Fig.~\ref{fig:validation}(a)]. At $J=0.3$ in the trivial phase, it
approaches a finite neutral excitation energy, $\widehat D_{p,M_0}(0)=0.364$,
$0.340$ and $0.330$ at $L=16$, $24$ and $32$, so
$L\widehat D_{p,M_0}(0)$ grows with $L$. At $J=1.5$, the lowest
opposite-parity state instead lies in the adjacent phase-winding sector, and
$L|\widehat D_{p,M_0}(0)|$ approaches $\pi^2\rho_s/2$. The size-dependent
zero $J_0(L)$ of this estimator separates the two behaviors. Centering each
curve on $J_0(L)$ produces a collapse versus $[J-J_0(L)]L$, consistent with
Ising finite-size scaling with $\nu=1$; the size dependence of $J_0(L)$ is
given in \SM{}~S3. The sequence of zeros moves toward the transition with
increasing $L$, but it is a finite-size consistency check rather than a
precision estimate of $J_c$. Applying the adjacent-winding protocol at
$J=0.3$ gives the corresponding control: the splitting keeps the same sign at
all nine winding--size combinations and at every flux sampled, so neither the
fixed-winding nor the adjacent-winding sign index changes, in contrast to the
topological phase. The $\pi$ shift is nonetheless still present there---the
symmetric combination of the $M_0\pm1$ responses reproduces the mirrored
central curve to a few percent---confirming that the identity itself does not
distinguish the phases, and that the branch exchange requires the flux
dispersion which only the topological phase supplies (\SM{}~S5).

Finite-entanglement scaling locates the critical region near
$J_c\simeq0.77$ using
$S=(c/6)\ln\xi_\chi+\mathrm{const}$~\cite{PollmannFES2009}
[Fig.~\ref{fig:validation}(b)]. The sequence $c\simeq1,3/2,1$ supports a
gapless charge mode and a continuous transition of the neutral Ising sector
over the accessible scales. In particular, $c\simeq1$ in both gapped neutral
phases is consistent with the remaining charge mode, while the increase to
$c\simeq3/2$ supports an additional critical Majorana mode. The central
charge supports the proposed critical degrees of freedom, while the
$L\widehat D_{p,M_0}$ scaling and winding-dependent parity response
distinguish the two phases. An open-chain calculation corroborates that
assignment: at $J=1.5$ the coherent particle-addition weight is confined to
the first few sites of the chain, whereas a trivial comparison spreads into
the bulk (\SM{}~S6). That calculation resolves the boundary localization
rather than the exponentially small end-to-end splitting, which stays below
the smooth background of the finite-difference estimators.
Together, the parity and entanglement scaling identify a continuous transition of the neutral Ising sector on top of a gapless charge mode.
Fit details and numerical limitations are given in \SM{}~S4; the
fixed-pairing Bogoliubov--de Gennes reference and the range of validity of
these statements are given in \SM{}~S7.

\emph{Experimental considerations.}---%
Assuming that the ring supports a stable one-dimensional paired regime, a
possible measurement would compare two metastable coplanar textures with adjacent
winding numbers in the same ring. In a helimagnetic ring, or a synthetic
platform with a programmable spin texture, flux-dependent spectroscopy of
Coulomb-blockade transitions could determine the charging-corrected
even--odd addition-energy splitting for each texture
~\cite{Yamada2025,BrauneckerSimon2013,KlinovajaStanoLoss2013,
Lafarge1993,Albrecht2016,LiuColeSau2019}. The predicted signal is the opposite leading flux
slope of the signed splitting for $M$ and $M+1$. Repeating the flux sweep for
both textures reduces common charging offsets, while texture-dependent
changes must be controlled. A symmetric comparison of textures on either side
of a reference winding would further reduce the leading smooth pitch
dependence, following the same construction used in the simulations.

\emph{Discussion.}---%
A local spin rotation maps a coplanar spiral onto a uniform Hamiltonian, but
on a ring it leaves the boundary sign $h_M=(-1)^M$. A single-valued combined
spin rotation and $U(1)$ gauge transformation converts this sign into the
exact boundary-phase shift $\pi M$ in every fixed-$N$ many-body spectrum. In
the topological phase, the global parity constraint assigns opposite fermion
parities to neighboring phase-winding sectors, so changing the texture
winding by one reverses their parity assignment. The topologically trivial
phase instead retains a finite
neutral gap and shows no corresponding reversal. For adjacent physical
textures, the exact boundary shift is accompanied by a
$2\pi/L$ pitch change, and the absolute parity remains a nonuniversal
microscopic property. The same boundary sign applies to nonuniform coplanar
textures with the same total winding. Thus the global fermion-parity response
is not determined by the local band mapping alone. Our results therefore identify magnetic-texture winding as a closed-geometry, number-conserving probe of topological pairing, complementary to diagnostics based on boundary modes or an externally imposed superconducting phase. More broadly, they show how the global winding of a magnetic texture can be encoded in the observable structure of an interacting many-body spectrum.

\emph{Acknowledgments.}---%
H.-Y.L. thanks Moon Jip Park for fruitful discussions. H.-Y.L. was
supported by the Basic Science Research Program through the National
Research Foundation of Korea funded by the Ministry of Science and ICT
[Grant Nos. RS-2023-00220471 and RS-2025-16064392].

\bibliography{refs}

\clearpage
\onecolumngrid

\begin{center}
{\large\bfseries Supplemental Material for\\[3pt]
``Magnetic-texture winding controls fermion-parity switches in an
interacting $p$-wave magnet ring''}
\end{center}

\setcounter{secnumdepth}{2}
\renewcommand{\thesection}{S\arabic{section}}
\renewcommand{\theHsection}{S\arabic{section}}
\setcounter{section}{0}
\renewcommand{\theequation}{S\arabic{equation}}
\renewcommand{\theHequation}{S\arabic{equation}}
\setcounter{equation}{0}
\renewcommand{\thefigure}{S\arabic{figure}}
\renewcommand{\theHfigure}{S\arabic{figure}}
\setcounter{figure}{0}
\renewcommand{\thetable}{S\arabic{table}}
\renewcommand{\theHtable}{S\arabic{table}}
\setcounter{table}{0}

This Supplemental Material supplies the derivations and numerical details
underlying the main text. Section~\ref{sm:exact} treats the exact spin
rotation, boundary condition, adjacent-texture correction, and fixed-$N$
spectra; Secs.~\ref{sm:lowenergy} and \ref{sm:parity} give the charge--Ising
theory, global parity constraint, and addition-energy estimators.
Sections~\ref{sm:methods} and \ref{sm:winding} present the numerical
procedures, finite-entanglement analysis, adjacent-winding data, and stiffness
comparison. Section~\ref{sm:edge} gives the open-chain particle-addition
profile and the smooth finite-difference backgrounds, and
Sec.~\ref{sm:scope} gives the fixed-pairing reference and states the scope of
the conclusions.

\section{Exact spin rotation and the winding-dependent boundary condition}
\label{sm:exact}

\subsection{Local rotation and uniform bulk Hamiltonian}

We label the sites of the ring by $j=0,\ldots,L-1$, set the lattice spacing
to one, and write the coplanar texture as
\begin{equation}
 {\bm n}_{M,j}
 =\bigl(\sin Q_Mj,0,\cos Q_Mj\bigr),\qquad
 Q_M=\frac{2\pi M}{L},\qquad M\in\mathbb Z .
\label{eq:sm-texture}
\end{equation}
The integer $M$ counts the number of turns made by the classical texture in
one circuit of the ring. With periodic physical electrons,
$c_{j+L}=c_j$, the Hamiltonian is
\begin{align}
H_M(\phi)={}&-t\sum_{j=0}^{L-1}
 \left(e^{i\phi/L}c_j^\dagger c_{j+1}+\mathrm{H.c.}\right)
\nonumber\\
&+J\sum_{j=0}^{L-1}c_j^\dagger
 \bigl[\sin(Q_Mj)\sigma^x+\cos(Q_Mj)\sigma^z\bigr]c_j
\nonumber\\
&-U\sum_{j=0}^{L-1}n_{j\uparrow}n_{j\downarrow}.
\label{eq:sm-model}
\end{align}
Here and below $\phi$ is the total Aharonov--Bohm phase acquired by a
single electron around the ring; $\phi=2\pi$ is one $h/e$ flux period.

Introduce the on-site spin rotation
\begin{equation}
 U_{M,j}=e^{-iQ_Mj\sigma^y/2},\qquad
 d_j=U_{M,j}^\dagger c_j .
\label{eq:sm-rotation}
\end{equation}
It aligns the exchange field,
\begin{equation}
U_{M,j}^\dagger
\bigl[\sin(Q_Mj)\sigma^x+\cos(Q_Mj)\sigma^z\bigr]
U_{M,j}=\sigma^z ,
\label{eq:sm-align}
\end{equation}
and gives the same link matrix on every interior bond,
\begin{equation}
G_M\equiv U_{M,j}^\dagger U_{M,j+1}
=e^{-iQ_M\sigma^y/2}.
\label{eq:sm-link}
\end{equation}
The on-site attraction is unchanged because
$n_{j\uparrow}n_{j\downarrow}=[n_j^2-n_j]/2$, with
$n_j=n_{j\uparrow}+n_{j\downarrow}$, is invariant under an on-site
spin rotation. Thus the transformation is exact for the interacting
Hamiltonian, rather than only at the single-particle level.

For a translationally invariant chain, and with the Fourier convention used
in the main text, the one-body matrix in the rotated frame is
\begin{equation}
\mathcal H_Q(k)
=-2t\cos\frac Q2\cos k\,\openone
+2t\sin\frac Q2\sin k\,\sigma^y
+J\sigma^z .
\label{eq:sm-band}
\end{equation}
The sign of the term proportional to $\sigma^y$ changes if the Fourier or
spiral-orientation convention is reversed, without changing its spectrum.
At $J=0$ the two eigenvalues are
$-2t\cos(k\pm Q/2)$, so the apparent odd-in-momentum splitting is only a
spin-dependent momentum shift. It becomes physically consequential together
with $J\sigma^z$, because the two spin matrices do not commute. Removing the
momentum shift then restores the spatially rotating exchange field.

\subsection{Closing the spin frame on a ring}
\label{sm:exact:boundary}

The local transformation does not by itself specify the boundary condition
of $d_j$. After one circuit,
\begin{equation}
U_{M,j+L}
=e^{-i\pi M\sigma^y}U_{M,j}
=(-1)^M U_{M,j},
\label{eq:sm-frameclose}
\end{equation}
and periodicity of $c_j$ therefore gives
\begin{equation}
d_{j+L}=(-1)^M d_j.
\label{eq:sm-dbc}
\end{equation}
The sign in Eq.~\eqref{eq:sm-dbc} is the center element of $SU(2)$ produced
by rotating a spinor through $2\pi M$. The classical moments are periodic for
every integer $M$, but the electronic spin frame is periodic only for even
$M$.

The same sign can be expressed as an ordinary boundary phase without using a
non-single-valued spin frame. Define
\begin{equation}
V_{M,j}=e^{i\pi Mj/L}U_{M,j},\qquad c_j=V_{M,j}f_j .
\label{eq:sm-V}
\end{equation}
Its two factors acquire the same sign after one circuit, and hence
\begin{equation}
V_{M,j+L}=e^{i\pi M}(-1)^M V_{M,j}=V_{M,j}.
\label{eq:sm-Vperiodic}
\end{equation}
The transformation is consequently a well-defined, particle-number
conserving unitary on the physical ring. On every bond it produces
\begin{equation}
e^{i\phi/L}V_{M,j}^\dagger V_{M,j+1}
=e^{i(\phi+\pi M)/L}e^{-iQ_M\sigma^y/2}.
\label{eq:sm-shiftedlink}
\end{equation}
Let $H_{\rm uni}(Q,\phi_{\rm eff})$ denote the Hamiltonian with this uniform link,
uniform exchange field $J\sigma^z$, periodic $f_j$, pitch parameter $Q$,
and effective boundary phase $\phi_{\rm eff}$.
Equations~\eqref{eq:sm-V}--\eqref{eq:sm-shiftedlink}
then prove
\begin{equation}
\operatorname{Spec}_N H_M(\phi)
=\operatorname{Spec}_N H_{\rm uni}\bigl(Q_M,\phi+\pi M\bigr),
\qquad \phi+\pi M\ \ {\rm mod}\ 2\pi .
\label{eq:sm-spectral}
\end{equation}
$\operatorname{Spec}_N$ denotes the spectrum in the fixed-$N$ sector. This
relation holds for every eigenvalue in every fixed-$N$ block, at any
$L$, $U$, $J$, and $\phi$. It does not assume a paired phase, a
long-wavelength limit, or a mean-field Hamiltonian. The parity of $M$
therefore changes the electronic boundary phase by $\pi$ even when the
many-body state is topologically trivial. Which fermion parity is lowest for
that boundary condition is a separate spectral question addressed in
Sec.~\ref{sm:parity}.

\subsection{Adjacent physical textures and symmetric pitch cancellation}
\label{sm:exact:adjacent}

Equation~\eqref{eq:sm-spectral} compares a physical texture to its uniform
rotated-frame image at the \emph{same} pitch. Comparing two physical textures
with windings $M$ and $M+1$ additionally changes the pitch by
\begin{equation}
\delta Q=Q_{M+1}-Q_M=\frac{2\pi}{L}.
\label{eq:sm-deltaQ}
\end{equation}
Writing $E_N^{\rm uni}(Q,\phi_{\rm eff})$ for an arbitrary fixed-$N$ energy level of
the uniform Hamiltonian, the exact relation gives
\begin{align}
E_{N,M+1}(\phi)
 &=E_N^{\rm uni}
 \left(Q_M+\delta Q,\phi+\pi M+\pi\right),\nonumber\\
E_{N,M}(\phi+\pi)
 &=E_N^{\rm uni}
 \left(Q_M,\phi+\pi M+\pi\right).
\label{eq:sm-adjacentexact}
\end{align}
The two expressions have exactly the same boundary phase and differ only in
their pitch argument. Thus the $\pi$ boundary-phase shift is exact, whereas
equality of the responses of adjacent physical textures is only approached
as $\delta Q\to0$.

This distinction also fixes the useful symmetric comparison. Let
$R(Q,\phi_{\rm eff})$ be any scaled response that is smooth in $Q$ within the same
phase; in particular, below we use
$R=L\widehat D_p$. About a reference winding $M_0$,
\begin{align}
\frac{R(Q_{M_0-\!1},\phi_{\rm eff})+R(Q_{M_0+\!1},\phi_{\rm eff})}{2}
&=R(Q_{M_0},\phi_{\rm eff})
+\frac{\delta Q^2}{2}\,\partial_Q^2R(Q_{M_0},\phi_{\rm eff})
+O(\delta Q^4).
\label{eq:sm-symmetric}
\end{align}
For the scaled response $R$, the pitch correction from either neighboring
texture is generically
$O(1/L)$, while the leading symmetric residual is $O(1/L^2)$. No exponent is
extracted from the four available sizes; Eq.~\eqref{eq:sm-symmetric} is used
only to identify which finite-size contribution is removed.

The boundary sign does not require a uniform spiral. For a nonuniform
coplanar texture described by an angle $\vartheta_j$ with
$\vartheta_{j+L}-\vartheta_j=2\pi M$, the rotation
$U_j=e^{-i\vartheta_j\sigma^y/2}$ still obeys
$U_{j+L}=(-1)^M U_j$. The rotated link matrices are then nonuniform, but the
combined transformation remains single valued:
\begin{equation}
V_j=e^{i\pi Mj/L}e^{-i\vartheta_j\sigma^y/2},
\qquad V_{j+L}=V_j .
\label{eq:sm-nonuniformV}
\end{equation}
It converts the global sign into the same phase shift $\pi M$, although the
rotated links retain the nonuniform increments
$\vartheta_{j+1}-\vartheta_j$. Local energies can therefore depend on the
detailed profile. For a noncoplanar texture the
accumulated spin rotation need not be a scalar center element, and a simple
$U(1)$ phase shift need not result.

\subsection{Complete fixed-$N$ spectra from exact diagonalization}
\label{sm:exact:ed}

We verified Eq.~\eqref{eq:sm-spectral} independently of the tensor-network
implementation by constructing the many-body Hamiltonian in occupation-number
bases. The comparison uses an interacting $L=6$ ring with $U=0.9$,
$J=0.73$, $M=1,2$, $\phi/\pi=0.37,0.63$, and $N=2,3,4$. The largest block
has dimension $\binom{12}{4}=495$. Within each block, all eigenvalues were
obtained and the sorted spectra were compared. The largest difference between
the laboratory-frame spectrum and the uniform-frame spectrum at the shifted
boundary phase was $5.8\times10^{-14}$. The agreement is not restricted to
the ground state.
The associated one-body unitary residuals and an independent
Jordan--Wigner implementation give the values summarized in
Table~\ref{tab:sm-ed}.

\par\medskip
\refstepcounter{table}\label{tab:sm-ed}
\begin{center}
\centering
\small
\begin{tabular}{p{0.42\textwidth}p{0.34\textwidth}c}
\hline\hline
Comparison & Scope & Maximum residual \\
\hline
Laboratory and locally rotated one-body matrices
& ring including the closing bond & $2.2\times10^{-16}$ \\
Rotated and uniform one-body matrices
& boundary phase shifted by $\pi M$ & $1.4\times10^{-15}$ \\
Complete interacting fixed-$N$ spectra
& $L=6$, $U=0.9$, $J=0.73$; $M=1,2$; $N=2,3,4$;
$\phi/\pi=0.37,0.63$ & $5.8\times10^{-14}$ \\
Tensor-network and Jordan--Wigner ground-state energies
& $L=4,5$, $M=1,2$, nonzero flux & $2.1$--$8.5\times10^{-14}$ \\
\hline\hline
\end{tabular}
\end{center}
\noindent\textbf{TABLE \thetable.}\quad
Numerical verification of the unitary relations. The spectral comparison
uses every eigenvalue in each indicated fixed-particle-number block;
matrix residuals use the operator norm.
\par\medskip

\section{Single-band pairing and the charge--Ising theory}
\label{sm:lowenergy}

\subsection{Projected odd-parity pairing channel}

The eigenvalues of Eq.~\eqref{eq:sm-band} are
\begin{equation}
\varepsilon_\pm(k)
=-2t\cos\frac Q2\cos k
\pm\sqrt{J^2+4t^2\sin^2\frac Q2\sin^2 k}.
\label{eq:sm-bands}
\end{equation}
At density $n=1/2$, reference pitch $Q=\pi/2$, and $J>0$, the Fermi level
crosses only the lower hybridized band. With one partially occupied band,
the one-dimensional particle count places its Fermi points at
$k_F=\pm\pi n=\pm\pi/2$. The single-band representation is exact for the
occupied noninteracting states once the upper band is empty. At $U=3$ it is
a low-energy organization rather than a parametrically controlled expansion,
because virtual pair excitations into the upper band are not governed by a
small ratio.

Let $u_-(k)$ be the lower-band spinor. Projecting an on-site singlet pair onto
that band gives
\begin{equation}
\Delta_-(k)
\propto u_-^{T}(k)i\sigma^y u_-(-k)
\propto
\frac{\alpha\sin k}{\sqrt{J^2+\alpha^2\sin^2 k}},
\qquad
\alpha=2t\sin\frac Q2 .
\label{eq:sm-projectedpair}
\end{equation}
The projected amplitude is odd under $k\to-k$ and vanishes at $k=0,\pi$.
This is the origin of the effective intraband $p$-wave pairing used in the
low-energy description. The microscopic attraction remains on-site and the
full Hamiltonian continues to conserve particle number.

\subsection{Gapless charge mode and neutral Majorana mass}

The number-conserving paired phase has a gapless charge mode and a gapped
neutral sector. Near a continuous transition between its trivial and
topological forms, the minimal continuum Hamiltonian is
\begin{equation}
H_{\rm low}=H_\rho+H_I+H_{\rho I}+\cdots ,
\label{eq:sm-lowtotal}
\end{equation}
with
\begin{align}
H_\rho&=\frac{u_\rho}{2\pi}\int_0^L dx\,
\left[
K_\rho(\partial_x\theta_\rho)^2+
K_\rho^{-1}(\partial_x\phi_\rho)^2
\right],\label{eq:sm-charge}\\
H_I&=\int_0^L dx\,
\left[
\frac{i v_I}{2}
\left(\chi_L\partial_x\chi_L-\chi_R\partial_x\chi_R\right)
+i m(J)\chi_L\chi_R
\right].
\label{eq:sm-ising}
\end{align}
The physical pair phase is
\begin{equation}
\Theta=\sqrt{2}\,\theta_\rho ,
\label{eq:sm-pairphase}
\end{equation}
and $\chi_{R,L}$ are the chiral components of the low-energy neutral
fermion. The sign of the Majorana mass $m(J)$ distinguishes the two gapped
neutral phases, while $|m|$ sets their long-distance neutral energy scale.
Additional massive neutral modes are absorbed into short-distance
parameters and into the nonuniversal parity offset introduced below.
This charge-boson plus Ising description is the number-conserving form of the
one-dimensional topological-pairing transition
~\cite{FLNF2011,RBA2015,KeselmanBerg2015,RuhmanAltman2017}.

Charge and Ising sectors need not be exactly decoupled in the lattice model.
A leading local coupling allowed by charge conservation has the form
\begin{equation}
H_{\rho I}
=g_{\rho I}\int_0^L dx\,
(\partial_x\phi_\rho)\,i\chi_L\chi_R .
\label{eq:sm-coupling}
\end{equation}
In this convention the smooth density fluctuation is
$\delta n=-(\sqrt{2}/\pi)\partial_x\phi_\rho$. The constant density component
is already absorbed into the Majorana mass $m(J)$.
Both factors have scaling dimension one at the decoupled critical point, so
Eq.~\eqref{eq:sm-coupling} is marginal by power counting, and density
fluctuations couple the gapless charge mode to the Ising energy density.
Such a coupling can modify the transition outside the weak-coupling regime
~\cite{Alberton2017}. We therefore use Eq.~\eqref{eq:sm-lowtotal} to identify
the candidate critical degrees of freedom and use the numerical scaling in
Sec.~\ref{sm:methods} to determine whether that description is realized for
$U=3$.

When $m\ne0$, the neutral sector is gapped and the charge boson gives central
charge $c=1$. If the continuous transition is governed by the decoupled
Ising fixed point, a critical Majorana mode contributes $c=1/2$, giving
\begin{equation}
c=1+\frac12=\frac32,\qquad \nu=1 .
\label{eq:sm-cnu}
\end{equation}
The central charge alone does not label which adjacent gapped phase is
topological. That information is supplied by the boundary-condition-dependent
parity spectrum derived next.

\subsection{Charge stiffness, compressibility, and $K_\rho$}
\label{sm:lowenergy:chargeparameters}

The charge Hamiltonian in Eq.~\eqref{eq:sm-charge} fixes two long-wavelength
responses,
\begin{equation}
\rho_s=\frac{2u_\rho K_\rho}{\pi},\qquad
D_2=L\,\partial_N^2E_{\rm sm}
     =\frac{\pi u_\rho}{2K_\rho},
\quad
u_\rho=\sqrt{\rho_sD_2},\quad
K_\rho=\frac{\pi}{2}\sqrt{\frac{\rho_s}{D_2}} .
\label{eq:sm-chargeweb}
\end{equation}
Here $D_2$ is the curvature of the smooth charging energy, with the
even--odd staggering removed. At $J=1.5$, the ring stiffness and charging
curvature give
\begin{center}
\begin{tabular}{ccccc}
\hline\hline
$L$ & $\rho_s$ & $D_2$ & $u_\rho$ & $K_\rho$ \\
\hline
24 & 0.5452 & 3.2201 & 1.325 & 0.646 \\
32 & 0.5461 & 3.2389 & 1.330 & 0.645 \\
40 & 0.5455 & 3.2476 & 1.331 & 0.644 \\
48 & 0.5460 & 3.2523 & 1.333 & 0.644 \\
\hline\hline
\end{tabular}
\end{center}
Thus $u_\rho\simeq1.33$ and $K_\rho\simeq0.644$ in the spinful
normalization of Eq.~\eqref{eq:sm-charge}. The low-energy spectrum contains
one occupied hybridized band, for which the corresponding single-channel
parameter is $\widetilde K=2K_\rho$. The noninteracting band gives
$\widetilde K=1$ from the same response definitions, whereas the interacting
value is $\widetilde K\simeq1.29$, consistent with the attractive
interaction.

An independent open-chain estimate follows from charge fluctuations across
a cut at position $x$,
\begin{equation}
\operatorname{Var}(N_A)
=\frac{K_\rho}{\pi^2}
\ln\!\left[\frac{2L}{\pi}\sin\frac{\pi x}{L}\right]+\mathrm{const}.
\label{eq:sm-chargefluctuation}
\end{equation}
The charge-resolved Schmidt values give $K_\rho=0.60(4)$. Their agreement
with Eq.~\eqref{eq:sm-chargeweb} supports a single gapless charge mode.
We do not use this comparison to infer a texture-pinned density harmonic or
to assign an additional instability.

\section{Global parity constraint, addition-energy estimator, and flux response}
\label{sm:parity}

\subsection{Neutral-fermion boundary condition and lowest-state parity}
\label{sm:parity:constraint}

In the paired low-energy sector, an electron in the occupied band can be
written schematically as a charge-$e$ vertex times a neutral fermion,
\begin{equation}
\psi_-(x)\sim e^{i\Theta(x)/2}\gamma(x).
\label{eq:sm-electronfactor}
\end{equation}
Let $W\in\mathbb Z$ be the winding of the pair phase,
\begin{equation}
\Theta(x+L)=\Theta(x)+2\pi W .
\label{eq:sm-W}
\end{equation}
The charge factor in Eq.~\eqref{eq:sm-electronfactor} then changes by
$(-1)^W$. In the rotated spin frame the electron also acquires the texture
sign $(-1)^M$. Single-valuedness of the physical electron fixes the neutral
boundary condition
\begin{equation}
\gamma(x+L)=\eta_\gamma\gamma(x),\qquad
\eta_\gamma=(-1)^{M+W}.
\label{eq:sm-neutralbc}
\end{equation}
Here $\eta_\gamma=+1$ and $-1$ are the periodic and antiperiodic neutral
boundary conditions, respectively. Equation~\eqref{eq:sm-neutralbc} fixes a
boundary condition; it does not by itself specify which neutral-fermion
parity has the lower energy.

The massive Ising spectrum supplies the second ingredient. In the
topological phase, the lowest states in the two neutral boundary conditions
have opposite fermion parities, up to a nonuniversal microscopic offset. In
the trivial phase they have the same relative parity, and the lowest state of
opposite parity costs a finite neutral excitation energy. This relative
statement follows directly from the massive-Majorana spectrum on a ring:
periodic (Ramond) momenta include the self-conjugate $k=0$ mode, whereas
antiperiodic (Neveu--Schwarz) momenta do not. A change in the sign of the
Majorana mass reverses the occupation of the lowest periodic-sector state and
therefore reverses its parity relative to the antiperiodic sector
~\cite{Kitaev2001,FLNF2011,RBA2015}. Ultraviolet modes can multiply both
assignments by a common sign but cannot change this relative inversion within
one connected phase. Since the gapless charge sector changes the particle
number in units of two, the total electron-number parity resides in the
neutral sector. Consequently, within a connected topological phase,
\begin{equation}
P_{\rm gs}(M,W)\equiv(-1)^{N_{\rm gs}(M,W)}
=\kappa\,(-1)^{M+W},
\qquad \kappa=\pm1 .
\label{eq:sm-globalparity}
\end{equation}
The sign $\kappa$ depends on microscopic filling and on ultraviolet neutral
modes; it is fixed as long as no additional parity crossing occurs within
the same phase. The universal statement is the relative one: changing $M$
or $W$ by one reverses the parity of the lowest state.

\begin{center}
\begin{tabular}{cccc}
\hline\hline
$(M+W)\bmod2$ & $\eta_\gamma$
& $P_{\rm gs}/\kappa$ (topological)
& $P_{\rm gs}/\kappa_{\rm triv}$ (trivial) \\
\hline
$0$ & $+1$ & $+1$ & $+1$ \\
$1$ & $-1$ & $-1$ & $+1$ \\
\hline\hline
\end{tabular}
\end{center}
Here $\kappa_{\rm triv}$ is another nonuniversal microscopic parity offset.
The table separates the exact boundary condition
$\eta_\gamma=(-1)^{M+W}$ from the phase-dependent parity of its lowest
state.

The energetic distinction between the two phases is most transparent as a
minimization problem. To leading order, the charge mode gives one
phase-winding branch for each $W$,
\begin{equation}
E_{\rm wind}(W;\phi)
=\frac{\rho_s}{2L}(\phi-\pi W)^2 ,
\label{eq:sm-windenergy}
\end{equation}
where $\rho_s$ is the thermodynamic charge stiffness in the convention of
the main text. Let $\Delta_{\rm n}$ denote the lowest neutral excitation in
the opposite-parity sector at fixed neutral boundary condition. In
the topological phase, the lowest energy at fixed total parity $P$ is
\begin{equation}
E_P^{(M)}(\phi)
=\min_{W\in\mathbb Z}
\left[
E_{\rm wind}(W;\phi)
+\frac{1-P\kappa(-1)^{M+W}}{2}\Delta_{\rm n}
\right]
+O(e^{-L/\xi_I})+O(L^{-2}),
\label{eq:sm-topmin}
\end{equation}
with $\xi_I$ the neutral correlation length. When
$\pi^2\rho_s/(2L)<\Delta_{\rm n}$, the lowest opposite-parity state uses an
adjacent $W$ branch rather than that neutral excitation. In the
trivial phase the lowest-state neutral parity
$\kappa_{\rm triv}$ is independent of the two boundary conditions:
\begin{equation}
E_P^{(M)}(\phi)
=\min_{W\in\mathbb Z}
\left[
E_{\rm wind}(W;\phi)
+\frac{1-P\kappa_{\rm triv}}{2}\Delta_{\rm n}
\right]+\cdots .
\label{eq:sm-trivmin}
\end{equation}
Equations~\eqref{eq:sm-topmin} and \eqref{eq:sm-trivmin} explicitly minimize
over the winding and neutral alternatives. For sufficiently large
topological rings, the $O(1/L)$ winding state is lower than the $O(1)$
neutral excitation. In the trivial phase, changing $W$ does not remove the
neutral cost.

\subsection{Charging-corrected even--odd staggering and finite-difference estimators}
\label{sm:parity:estimator}

Let $E_M(N,\phi)$ be the ground-state energy at fixed particle number.
Near an even reference value $N_0=L/2$, we separate a smooth function of
$N$ from the alternating addition-energy component,
\begin{equation}
E_M(N,\phi)
=E_{{\rm sm},M}(N,\phi)
-\frac{(-1)^N}{2}D_p^{(M)}(N,\phi)+\cdots .
\label{eq:sm-stagger}
\end{equation}
This equation defines the physical signed staggering after removal of the
smooth chemical-potential and charging contribution. With this convention
$D_p>0$ means that the corrected even sector lies lower. Below,
$D_p^{(M)}(\phi)\equiv D_p^{(M)}(N_0,\phi)$ unless the particle-number
argument is shown explicitly. Throughout the main text, the shorthand
$\widehat D_{p,M}$ denotes the five-point quantity
$\widehat D_{p,M}^{(5)}$ defined below.

The five-point quantity plotted in the main text is
\begin{align}
\widehat D_{p,M}^{(5)}(\phi)
=\frac18\big[&
4E_M(N_0-1,\phi)+4E_M(N_0+1,\phi)
-E_M(N_0-2,\phi)\nonumber\\
&-6E_M(N_0,\phi)-E_M(N_0+2,\phi)
\big].
\label{eq:sm-fivepoint}
\end{align}
Equivalently,
$\widehat D_{p,M}^{(5)}=-\Delta_N^4E_M/8$, where
\begin{equation}
\Delta_N^4E_M
=E_{-2}-4E_{-1}+6E_0-4E_{+1}+E_{+2}.
\label{eq:sm-fourth}
\end{equation}
and $E_m\equiv E_M(N_0+m,\phi)$. The fourth difference annihilates every
smooth polynomial of degree at most three and transmits a constant
alternating component with unit coefficient. If the physical staggering
varies slowly with density, the stencil also averages that envelope. For a
smooth bulk energy $E_{\rm sm}=Le(N/L)+O(1)$,
\begin{equation}
\widehat D_{p,M}^{(5)}
=D_p^{(M)}
+\frac12\,\partial_N^2D_p^{(M)}+\cdots
-\frac18\,\partial_N^4E_{\rm sm}
+O(\partial_N^6E_{\rm sm}).
\label{eq:sm-fiveerror}
\end{equation}
All particle-number derivatives are evaluated at $N=N_0$ and fixed $\phi$.
For a topological staggering
$D_p^{(M)}(N,\phi)=d_M(N/L,\phi)/L$, both the envelope term and the leading
smooth-background term are $O(L^{-3})$. They are subleading to the $O(1/L)$
physical signal by two powers of $L$.

We also use a seven-point quantity to assess the sensitivity to the smooth
charging background:
\begin{align}
\widehat D_{p,M}^{(7)}(\phi)=\frac1{32}\big[
&E_{-3}-6E_{-2}+15E_{-1}-20E_0\nonumber\\
&+15E_{+1}-6E_{+2}+E_{+3}
\big].
\label{eq:sm-sevenpoint}
\end{align}
The overall sign in Eq.~\eqref{eq:sm-sevenpoint} is chosen so that both
estimators return the same $D_p$ convention. It annihilates smooth
polynomials of degree at most five and gives
\begin{equation}
\widehat D_{p,M}^{(7)}
=D_p^{(M)}
+\frac34\,\partial_N^2D_p^{(M)}+\cdots
+\frac1{32}\,\partial_N^6E_{\rm sm}+\cdots .
\label{eq:sm-sevenerror}
\end{equation}
Thus the seven-point expression reduces the generic smooth charging
background from $O(L^{-3})$ to $O(L^{-5})$. It does not remove the density
dependence of the physical staggering; for
$D_p^{(M)}(N,\phi)=d_M(N/L,\phi)/L$, the leading envelope correction remains
$O(L^{-3})$.

\subsection{Flux-dependent parity splitting and finite-size curvatures}
\label{sm:parity:flux}

In the topological phase, Eq.~\eqref{eq:sm-globalparity} assigns opposite
parities to neighboring $W$ branches. On $0\le\phi\le\pi$, subtracting the
two leading branches gives
\begin{equation}
D_p^{(M)}(\phi)
\simeq s_M\frac{\pi\rho_s(Q_M)}{L}
\left(\phi-\frac{\pi}{2}\right),
\qquad s_M=\pm1,\qquad s_{M+1}=-s_M .
\label{eq:sm-linear}
\end{equation}
The amplitude varies smoothly with pitch through $\rho_s(Q_M)$. Flux tunes
the relative energies of neighboring phase-winding branches, while the
texture-winding parity fixes which branch is assigned even or odd particle
number. In the numerical data the branches cross near $\phi=\pi/2$.
Particle-number conservation prevents them from hybridizing at a crossing,
but it neither requires a crossing nor fixes its position.

At finite $L$, let $E_{e/o}^{\rm corr}$ denote the parity-resolved branch
energies after subtracting the smooth particle-number-dependent background.
We parameterize their wells by
\begin{equation}
E_{e/o}^{\rm corr}(\phi,L)
=A_{e/o}(L)
+\frac{\rho_{e/o}(L)}{2L}
\bigl[\phi-\phi_{e/o}^{\min}\bigr]^2+\cdots .
\label{eq:sm-paritywells}
\end{equation}
The quantities $\rho_e(L)$ and $\rho_o(L)$ are finite-size curvatures and
need not agree. Both approach the thermodynamic stiffness $\rho_s$. For the
present data, the measured branch assignment calibrates $\kappa=-1$ in
Eq.~\eqref{eq:sm-globalparity}. Thus, for the even reference winding, the odd
minimum is at $\phi=0$ and the even minimum at $\phi=\pi$. If the
neutral-sector offset $A_o-A_e$ is exponentially small,
Eq.~\eqref{eq:sm-paritywells} and the estimator expansion give
\begin{equation}
\widehat D_{p,M_0}^{(5)}(0)
=-\frac{\pi^2\rho_e(L)}{2L}
+O(e^{-L/\xi_I})+O(L^{-3}),
\label{eq:sm-stiffnessrelation}
\end{equation}
where the last term includes higher harmonics, the density dependence of the
staggering envelope, and the five-point smooth-background correction. This
is a leading-harmonic finite-size relation, not an exact identity. The
numerical comparison is given in Sec.~\ref{sm:winding:stiffness}.

\subsection{Size-dependent zero and recentered finite-size collapse}
\label{sm:parity:J0}

The zero-flux estimator changes from a finite neutral cost in the trivial
phase to an $O(1/L)$ winding energy in the topological phase. For each size
we define $J_0(L)$ by
\begin{equation}
\widehat D_{p,M_0}^{(5)}(0;J_0,L)=0 .
\label{eq:sm-J0def}
\end{equation}
Linear interpolation between the two nearest calculated couplings gives
\begin{center}
\begin{tabular}{cccc}
\hline\hline
$L$ & Bracketing interval & $J_0(L)$ & $J_0(L)-0.774$ \\
\hline
16 & $0.82$--$0.84$ & $0.82187$ & $0.04787$ \\
24 & $0.78$--$0.80$ & $0.79909$ & $0.02509$ \\
32 & $0.78$--$0.80$ & $0.78919$ & $0.01519$ \\
\hline\hline
\end{tabular}
\end{center}
The sequence moves toward the critical region near $J\simeq0.77$, but these
three zeros are not used to obtain a precision thermodynamic critical
coupling. Their more limited role is to center the local finite-size
comparison in the inset of Fig.~3(a) of the main text.

Using
\begin{equation}
x=[J-J_0(L)]L
\label{eq:sm-collapsex}
\end{equation}
gives a collapse over the displayed common window $|x|\le1$. A joint linear
fit in that window gives
\begin{equation}
L\widehat D_{p,M_0}^{(5)}
=-0.780\,x-0.0018,\qquad
{\rm rms\ residual}=2.8\times10^{-3}.
\label{eq:sm-collapsefit}
\end{equation}
The factor of $L$ in Eq.~\eqref{eq:sm-collapsex} is consistent with the
Ising correlation-length exponent $\nu=1$. Because each curve is centered
on its own measured zero, this collapse probes the local scaling shape; it
does not independently determine $J_c$.

For comparison, the fixed-texture $J=0.2$, $L=24$ data give
\begin{equation}
L\widehat D_{p,M_0}^{(5)}(0)=+9.34,\qquad
L\widehat D_{p,M_0}^{(5)}(\pi)=+10.24 .
\label{eq:sm-trivialendpoints}
\end{equation}
The two endpoints have the same sign and therefore show no endpoint sign
reversal between $\phi=0$ and $\pi$. These data do not resolve the interior
flux dependence and do not compare adjacent texture windings; the
adjacent-winding calculations in Sec.~\ref{sm:winding} are at $J=1.5$.

\section{Numerical methods, convergence, and finite-entanglement scaling}
\label{sm:methods}

\subsection{Finite rings and exact-diagonalization benchmarks}

Finite-ring ground states were obtained with particle-number-conserving DMRG
as implemented in TeNPy~v1.0.6
~\cite{White1992,Schollwoeck2011,TeNPy}. The spiral has a transverse
$\sigma^x$ component, so no uniform spin component is conserved; the only
symmetry used in the matrix-product states is total charge $U(1)_N$. At zero
flux the laboratory-frame Hamiltonian is real. A nonzero Peierls phase makes
the hopping complex, and all flux-dependent calculations therefore use
complex tensors. Energies were obtained independently in each particle-number
sector entering Eqs.~\eqref{eq:sm-fivepoint} and
\eqref{eq:sm-sevenpoint}.

The largest bond dimensions and the principal independent comparisons are summarized in
Table~\ref{tab:sm-numerics}. The scaled splitting at $J=1.5$ was computed
with $\chi=800$ for $L\le32$, $\chi=1200$ for $L=40$, and $\chi=1600$ for
$L=48$. Repeating the $L=24$ and $32$ reference-winding calculations with
$\chi=1200$ changes $L\widehat D_p$ by less than the sixth displayed decimal.
For the $L=24$ flux curve, the shared $M=6$ points obtained in the
adjacent-winding calculation agree with the independently repeated
reference-winding calculation to four decimals in $L\widehat D_p$.

\begin{table}[t]
\centering
\small
\begin{tabular}{
p{0.20\textwidth}
p{0.30\textwidth}
p{0.14\textwidth}
p{0.27\textwidth}}
\hline\hline
Quantity & Geometry and parameters & Bond dimension & Convergence information \\
\hline
$J$ dependence of $L\widehat D_p$
& rings, $L=16,24,32$; additional $J=1.5$ points at $L=40,48$
& $800$--$1600$
& $\chi=1200$ repetitions at $L=24,32$ agree to six decimals \\
Reference-winding flux response
& rings, $L=16,20,24,28,32$
& $800$; selected $\chi=1200$
& five- and seven-point zero locations compared; representative endpoint repeated \\
Adjacent-winding endpoints
& $L=16,20,24,32$; $M_0-1,M_0,M_0+1$; $\phi=0,\pi$
& $800$
& observed final-sweep drift propagated through each finite difference \\
Adjacent-winding flux curves
& $L=24$; $M=5,6,7$; three interior and two endpoint flux values
& $800$
& all 45 constituent energies at the interior flux values obtained; one point repeated three times \\
Finite-entanglement scaling
& infinite chain, four-site unit cell
& $\chi=100$--$1131$; neighboring couplings followed to $1600$
& four inequivalent bond cuts; restricted fits use
$\chi\ge200$ or $\chi\ge283$ as specified below \\
Exact diagonalization
& $L=4$--$6$ rings and fixed-$N$ blocks
& full Hilbert blocks
& matrix and complete-spectrum residuals in Table~\ref{tab:sm-ed} \\
\hline\hline
\end{tabular}
\caption{\label{tab:sm-numerics}
Numerical calculations entering the main text. Absolute energies, their
finite differences, and fitted curvatures were monitored separately because
the relevant error of a signed addition-energy estimator is the propagated
error of its linear combination, rather than the absolute extensive energy.}
\end{table}

For the adjacent-winding endpoints, 115 of the 120 energies entering the
finite differences satisfied the convergence criterion, and all five
energies did so for 19 of the 24 endpoint values. For the remaining cases,
propagating the observed final-sweep energy drift with the weights in
Eq.~\eqref{eq:sm-fivepoint} gives at most $3.8\times10^{-7}$ in
$\widehat D_p$. This quantity estimates the observed numerical drift and is
not a rigorous error bound. The smallest endpoint magnitude used to determine
a sign is approximately $7.8\times10^{-2}$, more than five orders of
magnitude larger. We therefore use these data to determine the signs and the
reduction produced by symmetric averaging, but do not infer a thermodynamic
exponent from their amplitudes.

One $L=24$ calculation, at
$(M,N,\phi)=(7,10,3\pi/4)$, did not satisfy the convergence criterion within
the sweeps performed. Three independent repetitions give an energy spread of
$1.3\times10^{-8}$. Separately propagating the recorded final-sweep drifts
of all five energies entering the estimator gives
$8.4\times10^{-8}$ in $\widehat D_p$, compared with
$|\widehat D_p|=5.05\times10^{-2}$.

Small-ring exact diagonalization serves two distinct purposes. The complete
spectra in Sec.~\ref{sm:exact:ed} verify the implementation of the exact
unitary mapping, while direct DMRG--ED comparisons at the same Hamiltonian
parameters verify the flux and particle-number conventions. These
small-system comparisons do not replace bond-dimension convergence at the
larger sizes.

\subsection{Finite-entanglement scaling}
\label{sm:methods:fes}

The infinite-system calculations use a four-site unit cell, matching the
reference texture, and preserve $U(1)_N$. At finite bond dimension $\chi$, an
infinite matrix-product state has a finite transfer-matrix correlation length
\begin{equation}
\xi_\chi=-\frac{\ell_{\rm cell}}{\ln|\lambda_2|},
\qquad \ell_{\rm cell}=4 ,
\label{eq:sm-xi}
\end{equation}
where $\lambda_2$ is the subleading transfer eigenvalue in the
zero-$U(1)$-charge sector of the transfer matrix, used consistently for the
three curves in Fig.~3(b) of the main text. This transfer-matrix sector can
contain both density and neutral operators. We therefore use $\xi_\chi$ to
extract the total effective central charge and do not identify an individual
transfer eigenvalue with a neutral gap.

At a conformal point, the entanglement entropy at an inequivalent bond cut
obeys
\begin{equation}
S_\chi=\frac{c}{6}\ln\xi_\chi+s_0+o(1).
\label{eq:sm-fes}
\end{equation}
Entropies on all four bonds of the unit cell were recorded. For visual
clarity, the main-text figure shows a maximal-entropy bond cut at each
coupling; the four-cut analysis below is used for the quantitative
comparison. Table~\ref{tab:sm-fes} gives ordinary least-squares slopes for
all eight displayed points ($\chi=100$--$1131$) and for the six points with
the largest $\xi_\chi$ ($\chi\ge200$). The difference between the two
columns is a fit-window estimate of finite-$\chi$ drift, not a statistical
standard error.

\begin{table}[t]
\centering
\begin{tabular}{cccc}
\hline\hline
$J$ & CFT reference value & $c$ (all eight points) & $c$ (largest six $\xi_\chi$) \\
\hline
$0.2$ & $1$ & $0.949$ & $0.959$ \\
$0.774$ & $3/2$ & $1.426$ & $1.469$ \\
$1.5$ & $1$ & $0.992$ & $0.986$ \\
\hline\hline
\end{tabular}
\caption{\label{tab:sm-fes}
Finite-entanglement slopes of the representative curves plotted in
Fig.~3(b) of the main text. The increase of the critical slope as the
smallest-$\xi_\chi$ points are removed is consistent with approach to
$c=3/2$.}
\end{table}

The four inequivalent cuts give the same phase sequence. Fitting the five
points with $\chi\ge283$ at every cut gives bond-averaged values
$c=0.978$, $1.522$, and $0.985$ at $J=0.2$, $0.774$, and $1.5$,
respectively. At $J=0.774$ the individual values are
$1.467,1.467,1.577,1.577$; this bond-to-bond spread is the dominant
systematic uncertainty and explains why the representative curve in
Table~\ref{tab:sm-fes} has a lower slope than the four-bond average.
At neighboring couplings, the running slopes bend below the $c=3/2$ line at
$J=0.7725$ and above it at $J=0.775$ as $\chi$ increases. We therefore use
the $J=0.774$ curve as representative of the critical region near
$J_c\simeq0.77$, not as an independent high-precision estimate of $J_c$.
Taken together with the parity-splitting scaling, the sequence
$c\simeq1,3/2,1$ supports, over the accessible bond dimensions, a continuous
transition with one gapless charge boson and one critical Majorana mode.

\section{Adjacent-winding response and stiffness comparison}
\label{sm:winding}

\subsection{Endpoint alternation at four ring sizes}
\label{sm:winding:endpoints}

For a reference winding $M_0=L/4$, define the signed endpoint amplitude
\begin{equation}
A_M=\frac{L}{2}\left[
\widehat D_{p,M}^{(5)}(\pi)-\widehat D_{p,M}^{(5)}(0)
\right].
\label{eq:sm-amplitude}
\end{equation}
Its sign is the leading flux slope. Table~\ref{tab:sm-endpoints} collects
the three adjacent windings for all four sizes. The sign of $A_M$ alternates
within every triplet, showing the reversal of the leading flux slope
under $M\to M+1$. The absolute offset is size dependent through $M_0$, but
the relative alternation within each triplet is independent of that offset.

\begin{table}[t]
\centering
\begin{tabular}{ccc}
\hline\hline
$L$ & $(M_-,M_0,M_+)$ & $(A_-,A_0,A_+)$ \\
\hline
16 & $(3,4,5)$ & $(-2.599,\,+2.668,\,-1.964)$
\\
20 & $(4,5,6)$ & $(+2.789,\,-2.680,\,+2.293)$
\\
24 & $(5,6,7)$ & $(-2.894,\,+2.686,\,-2.424)$
\\
32 & $(7,8,9)$ & $(-2.853,\,+2.692,\,-2.514)$
\\
\hline\hline
\end{tabular}
\caption{\label{tab:sm-endpoints}
Endpoint amplitudes at $J=1.5$. Their alternating signs encode the reversal
of the leading flux slope. The estimator is Eq.~\eqref{eq:sm-fivepoint};
convergence information is given in Sec.~\ref{sm:methods}.}
\end{table}

\subsection{Full $L=24$ curves and cancellation of the pitch correction}
\label{sm:winding:curves}

The five flux values used in Fig.~2(a) of the main text are listed in
Table~\ref{tab:sm-curves}. They show directly that the $M=6$ curve has the
opposite slope from the $M=5$ and $M=7$ curves. Linear interpolation between
the two values nearest zero gives
\begin{equation}
\phi_*/\pi=0.5018,\quad0.4985,\quad0.5031
\qquad (M=5,6,7),
\label{eq:sm-crossings}
\end{equation}
respectively. The grid spacing, smooth charging residual, and finite-size
pitch shift set the significance of these numbers; they are not used as
precision crossing positions.

\begin{table}[t]
\centering
\begin{tabular}{cccc}
\hline\hline
$\phi/\pi$ & $L\widehat D_{p,5}^{(5)}$
& $L\widehat D_{p,6}^{(5)}$ & $L\widehat D_{p,7}^{(5)}$ \\
\hline
$0$   & $+2.9040$ & $-2.6776$ & $+2.4655$ \\
$1/4$ & $+1.4580$ & $-1.3355$ & $+1.2413$ \\
$1/2$ & $+0.0103$ & $+0.0082$ & $+0.0151$ \\
$3/4$ & $-1.4374$ & $+1.3517$ & $-1.2116$ \\
$1$   & $-2.8835$ & $+2.6937$ & $-2.3822$ \\
\hline\hline
\end{tabular}
\caption{\label{tab:sm-curves}
Charging-corrected adjacent-winding responses at $J=1.5$ and $L=24$.}
\end{table}

The exact boundary shift and the smooth pitch change can be separated
pointwise. Define
\begin{equation}
R_M(\phi)=L\widehat D_{p,M}^{(5)}(\phi).
\label{eq:sm-R}
\end{equation}
For the real laboratory-frame Hamiltonian,
complex conjugation and $2\pi$ flux periodicity give
$R_{M_0}(\phi+\pi)=R_{M_0}(\pi-\phi)$ on
$0\le\phi\le\pi$. Equation~\eqref{eq:sm-adjacentexact} therefore motivates
\begin{align}
r_\pm(\phi)&=R_{M_0\pm1}(\phi)-R_{M_0}(\pi-\phi),\nonumber\\
r_{\rm sym}(\phi)&=
\frac{R_{M_0-1}(\phi)+R_{M_0+1}(\phi)}{2}
-R_{M_0}(\pi-\phi).
\label{eq:sm-residuals}
\end{align}
At $L=24$, the unnormalized root-mean-square residuals are $0.1472$,
$0.1827$, and $0.0211$ for $r_-$, $r_+$, and $r_{\rm sym}$,
respectively. Normalizing them by
$\sqrt{\langle R_{6}(\pi-\phi)^2\rangle}=1.8993$, where the average is over
the five flux values in Table~\ref{tab:sm-curves}, gives
$0.0775$, $0.0962$, and $0.0111$; the normalized maximum of the symmetric
residual is $0.0236$. The corrections from the two neighboring textures have
opposite signs away from the crossing, and averaging reduces their
root-mean-square size by
factors of $7.0$ and $8.7$. This is the full-curve version of the Taylor
cancellation in Eq.~\eqref{eq:sm-symmetric}.

Across the four sizes, define
\begin{equation}
C_L=\frac{\left|(A_-+A_+)/2+A_0\right|}{\overline{|A|}},
\qquad
S_L=\frac{|A_+-A_-|}{\overline{|A|}},
\label{eq:sm-CS}
\end{equation}
where $\overline{|A|}$ is the mean of the three amplitude magnitudes.
$S_L$ compares the two neighbors, which have the same boundary sign but
different pitches, while $C_L$ is the normalized residual after symmetrically
averaging the two neighboring responses. The values are
\begin{center}
\begin{tabular}{ccccc}
\hline\hline
$L$ & $16$ & $20$ & $24$ & $32$ \\
\hline
$C_L$ & $0.1604$ & $0.0536$ & $0.0101$ & $0.00315$ \\
$S_L$ & $0.2638$ & $0.1916$ & $0.1761$ & $0.1261$ \\
\hline\hline
\end{tabular}
\end{center}
The product $LS_L$ remains near four, consistent with a leading
$O(1/L)$ difference between the two neighboring textures. The faster
decrease of $C_L$ is consistent with cancellation of the term linear in
$\delta Q$. Four sizes are insufficient to assign an asymptotic exponent to
$C_L$.

\subsection{Parity-resolved curvature and the winding-energy scale}
\label{sm:winding:stiffness}

The even- and odd-particle-number energies in Fig.~2(b) of the main text are
referenced to the minimum within each sector before fitting. This removes the
smooth difference in chemical potential between $N=12$ and $N=13$ and keeps
only the flux dependence. Quadratic fits over
$|\phi-\phi_{e/o}^{\min}|\le\pi/2$, constrained to pass through the
corresponding sector minimum, give the finite-size curvatures in
Table~\ref{tab:sm-curvatures}. The dashed curves in main-text Fig.~2(b) use
the common guide $\rho_s=0.5461$ obtained from the large-size extrapolation
described below; the data retain the distinct finite-size curvatures listed
here.

\begin{table}[t]
\centering
\begin{tabular}{ccccc}
\hline\hline
$L$ & $\rho_e(L)$ & $\rho_o(L)$
& $\pi^2\rho_e(L)/2$ & $L|\widehat D_{p,M_0}^{(5)}(0)|$ \\
\hline
16 & $0.5506$ & $0.6129$ & $2.717$ & --- \\
24 & $0.5483$ & $0.5928$ & $2.706$ & $2.6776$ \\
32 & $0.5478$ & $0.5818$ & $2.703$ & $2.6874$ \\
\hline\hline
\end{tabular}
\caption{\label{tab:sm-curvatures}
Finite-size parity-sector curvatures at $J=1.5$. The two curvatures differ
at finite $L$ but move toward a common large-$L$ value. The final
column permits a same-size comparison with
Eq.~\eqref{eq:sm-stiffnessrelation}.}
\end{table}

The comparison relevant to Eq.~\eqref{eq:sm-stiffnessrelation} is made
without using the odd--even energy difference to determine the stiffness.
Using the common curvature window
$|\phi-\phi_{e/o}^{\min}|\le\pi/2$, we fit the even-sector values to
$\rho_e(L)=\rho_s+a/L$. Retaining $L=16,24,32$ gives
$\rho_s=0.5447$, while retaining the two largest sizes gives
$\rho_s=0.5461$. We use the latter for the dashed guide in the main figure
and regard the range $0.5447$--$0.5461$ as the finite-size extrapolation
variation. It corresponds to
$\pi^2\rho_s/2=2.688$--$2.695$.
The reference-winding zero-flux values are
\begin{center}
\begin{tabular}{cccccc}
\hline\hline
$L$ & $24$ & $32$ & $40$ & $48$ & bulk-curvature prediction \\
\hline
$L|\widehat D_{p,M_0}^{(5)}(0)|$
& $2.6776$ & $2.6874$ & $2.6919$ & $2.6943$
& $2.688$--$2.695$ \\
\hline\hline
\end{tabular}
\end{center}
Across this extrapolation range, the large-$L$ scaled splitting agrees with the
bulk-curvature prediction at the percent level. The more direct same-size
comparison in Table~\ref{tab:sm-curvatures} has additional corrections from
higher flux harmonics, the neutral-sector offset, and the estimator envelope
and smooth background. The agreement supports the interpretation of the
$O(1/L)$ parity splitting as the phase-winding energy. The curvature alone
remains a charge response and is not a topological diagnostic.

For the reference periodic ring at $L=24$,
$L\widehat D_p^{(5)}(0)=-2.678$ and
$L\widehat D_p^{(5)}(\pi)=+2.694$. The five-point and seven-point crossing
locations are $\phi_*^{(5)}/\pi=0.4985$ and
$\phi_*^{(7)}/\pi=0.5000$. The seven-point expression moves the small
five-point offset near $\phi=\pi/2$ toward zero. Its generic smooth charging
contribution is $O(L^{-5})$, while a density-dependent parity-staggering
envelope can still contribute at $O(L^{-3})$.

\subsection{Trivial-phase control at $J=0.3$}
\label{sm:winding:trivial}

\begin{figure}[H]
\centering
\includegraphics[width=0.6\textwidth]{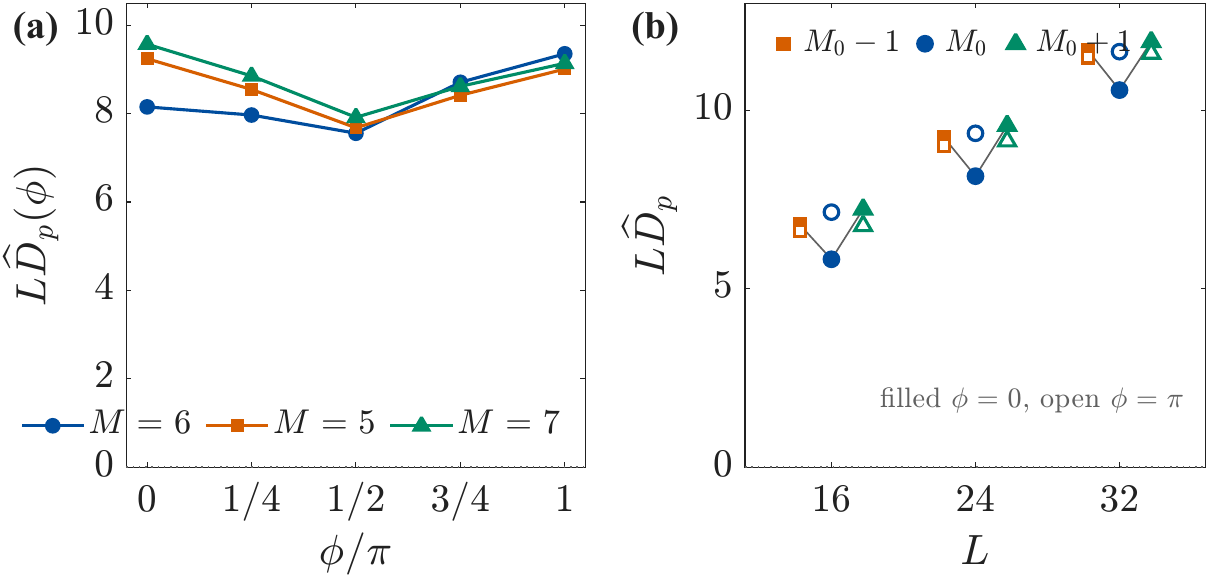}
\caption{\label{fig:sm-trivial}
Adjacent-winding response in the trivial phase ($J=0.3$, $U=3$), obtained with
the protocol used at $J=1.5$. (a) Scaled splitting
$L\widehat D_{p,M}(\phi)$ at $L=24$ for $M=5,6,7$. The three curves stay
positive and track one another; compare Fig.~2(a) of the main text, where they
alternate in sign and cross zero near $\phi=\pi/2$. (b) Endpoint values at
$L=16,24,32$ for the triplet $M_0-1$, $M_0$, $M_0+1$ (filled $\phi=0$, open
$\phi=\pi$). No sign changes with winding or with flux, and the scaled
splitting grows with $L$, as expected for a finite neutral gap.}
\end{figure}

The exact boundary-phase shift of Eq.~\eqref{eq:sm-spectral} is
independent of the many-body phase, so the adjacent-winding protocol was
repeated in the trivial phase with the settings used at $J=1.5$
($\chi=800$, $50$ sweeps, energy tolerance $10^{-10}$, boundary flux gauge).
The coupling is $J=0.3$, the trivial reference point of the finite-ring
$J$ scan; $J=0$ is not usable, because the texture coupling itself vanishes
there and no winding can be defined.

Table~\ref{tab:sm-trivial-endpoints} and Fig.~\ref{fig:sm-trivial}(b) give
the endpoint values. The
splitting is positive at every winding and every size, so the even sector
lies lower throughout: the fixed-winding index $\nu_M$ is $+1$ in all nine
cases and the adjacent-winding index is $+1$ in all six pairs. Neither index
changes, in contrast to the topological phase, where both alternate in every
case examined. The magnitude is set by a finite neutral excitation energy
rather than by a phase-winding energy: $\widehat D_{p,M_0}(0)=0.364$, $0.340$
and $0.330$ at $L=16$, $24$ and $32$, so the scaled quantity
$L\widehat D_{p,M_0}(0)$ grows with $L$ instead of saturating.

\begin{center}
\begin{tabular}{cccc}
\hline\hline
$L$ & $M$ & $L\widehat D_{p,M}(0)$ & $L\widehat D_{p,M}(\pi)$ \\
\hline
$16$ & $3$ & $6.794$ & $6.619$ \\
$16$ & $4$ & $5.824$ & $7.144$ \\
$16$ & $5$ & $7.228$ & $6.762$ \\
$24$ & $5$ & $9.241$ & $9.008$ \\
$24$ & $6$ & $8.152$ & $9.353$ \\
$24$ & $7$ & $9.570$ & $9.139$ \\
$32$ & $7$ & $11.679$ & $11.482$ \\
$32$ & $8$ & $10.564$ & $11.640$ \\
$32$ & $9$ & $11.918$ & $11.591$ \\
\hline\hline
\end{tabular}
\end{center}
\noindent\textbf{TABLE \thetable.}\quad
\refstepcounter{table}\label{tab:sm-trivial-endpoints}%
Trivial-phase endpoints at $J=0.3$. All entries are positive, so the sign
indices do not change with winding or with flux.

\medskip
The full five-flux curves at $L=24$ [Fig.~\ref{fig:sm-trivial}(a)] show the
same behaviour across the interval, and also show that the $\pi$ shift itself is still present. Writing
$R_M(\phi)=L\widehat D_{p,M}(\phi)$, the symmetric combination
$\tfrac12[R_{M_0-1}(\phi)+R_{M_0+1}(\phi)]$ reproduces the mirrored central
curve $R_{M_0}(\pi-\phi)$ to $0.6\%$, $0.1\%$, $3.2\%$, $6.9\%$ and $11.3\%$
at $\phi/\pi=0$, $1/4$, $1/2$, $3/4$ and $1$, a mean deviation of $4.4\%$
consistent with the $O(1/L)$ pitch detuning at this size. The identity
therefore holds on both sides of the transition, while the response it acts
on is nearly flat here: shifting it by $\pi$ does not change its sign.
\medskip
Provenance: $135$ ground-state energies, all completed. The sweep-to-sweep
energy change fell below the requested $10^{-10}$ in $94$ of them; the
largest energy uncertainty over the whole set is $2.4\times10^{-5}$, giving a
propagated uncertainty of at most $1.9\times10^{-5}$ in $\widehat D_{p,M}$.
That is four orders of magnitude below the smallest entry in
Table~\ref{tab:sm-trivial-endpoints} divided by $L$, so it affects neither
the signs nor the size scaling reported here.

\section{Open-chain particle-addition profile and finite-difference backgrounds}
\label{sm:edge}

\begin{figure}[H]
\centering
\includegraphics[width=0.6\columnwidth]{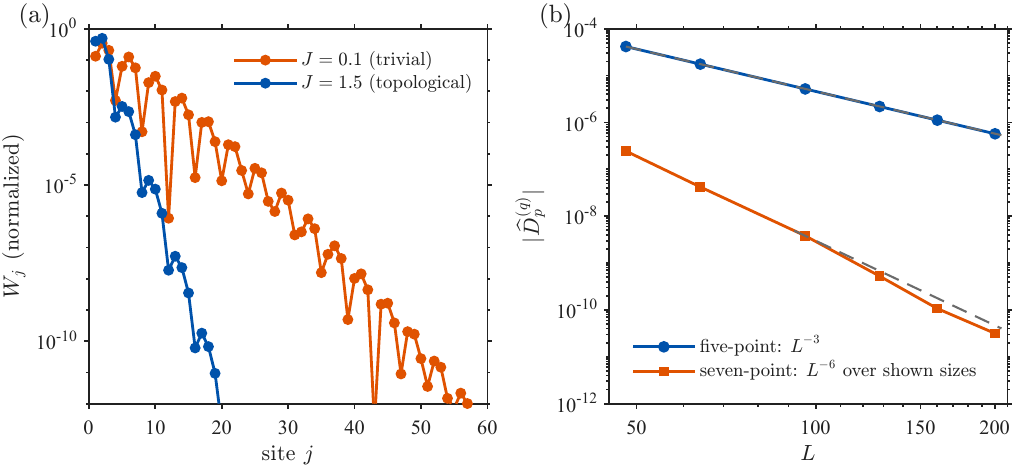}
\caption{\label{fig:sm-edge}
Open-chain results at $L=64$ in panel (a) and $J=1.5$ in panel (b).
(a)~Normalized coherent particle-addition weight in the topological regime
($J=1.5$) and a trivial comparison ($J=0.1$). The one-sided profile is one
state selected from the nearly degenerate edge subspace.
(b)~Absolute five- and seven-point estimators. The observed $L^{-3}$ and
$L^{-6}$ dependences are smooth finite-difference backgrounds, not a direct
measurement of the physical edge-state splitting.}
\end{figure}

\subsection{Edge-localized coherent particle-addition weight}

For open boundaries, we evaluate the coherent part of the lowest
particle-addition profile,
\begin{equation}
W_j=
\frac{\displaystyle\sum_\sigma
\left|\langle\Psi_{N_0+1}|c^\dagger_{j\sigma}|\Psi_{N_0}\rangle\right|^2}
{\displaystyle\sum_{\ell,\sigma}
\left|\langle\Psi_{N_0+1}|c^\dagger_{\ell\sigma}|\Psi_{N_0}\rangle\right|^2}.
\label{eq:sm-edgeweight}
\end{equation}
At $J=1.5$ and $L=64$, $99.4\%$ of this normalized weight lies within the
first four sites and $99.998\%$ within the first eight
[Fig.~\ref{fig:sm-edge}(a)]. The $J=0.1$ comparison retains a pronounced
Friedel pattern and extends farther into the chain. The contrast shows that
the coherent odd-particle addition is localized at the boundary in the
topological regime.

The displayed state is localized at one end. This does not assign a
preferred physical end to the wire: a finite-system solver may return one
linear combination from a nearly degenerate edge subspace. Establishing
separate left- and right-localized states would require resolving or jointly
targeting the two lowest states in the odd sector. We therefore use
Fig.~\ref{fig:sm-edge}(a) only as an edge-localization diagnostic, rather
than as a reconstruction of two individual Majorana wavefunctions.

\subsection{Smooth backgrounds in the five- and seven-point estimators}

The open-chain ground-state energy has the smooth finite-size expansion
\begin{equation}
E_{\rm sm}(N,L)
=L e(n)+a(n)+\frac{b(n)}{L}+\cdots,\qquad n=\frac NL ,
\label{eq:sm-openexpansion}
\end{equation}
in addition to the physical even--odd staggering. Applying the estimators
in Eqs.~\eqref{eq:sm-fivepoint} and \eqref{eq:sm-sevenpoint} gives
\begin{align}
\widehat D_p^{(5)}-D_p
&=-\frac18\left[
\frac{e^{(4)}(n)}{L^3}+\frac{a^{(4)}(n)}{L^4}+\cdots\right],
\nonumber\\
\widehat D_p^{(7)}-D_p
&=+\frac1{32}\left[
\frac{e^{(6)}(n)}{L^5}+\frac{a^{(6)}(n)}{L^6}+\cdots\right].
\label{eq:sm-openartifacts}
\end{align}
Terms from a slowly density-dependent staggering and higher derivatives are
not written. Equation~\eqref{eq:sm-openartifacts} shows that a central
difference can display an algebraic tail even when the physical edge-state
splitting lies below it.

For $J=1.5$ and $48\le L\le200$, the five-point values follow $L^{-3}$.
The compensated fourth derivative approaches
$L^3\partial_N^4E_{\rm sm}\simeq-36.6$ and accounts for the measured
five-point values within $0.5\%$, identifying this tail with the bulk
charging term. The seven-point values follow approximately $L^{-6}$ over
the same sizes. This is consistent with a small bulk coefficient
$e^{(6)}(1/2)$, so that the boundary term proportional to $a^{(6)}$ dominates
the accessible range.

The generic leading smooth background of the seven-point estimator remains
$L^{-5}$; the observed $L^{-6}$ regime therefore does not eliminate all
smooth backgrounds and need not persist to arbitrarily large $L$. The data in
Fig.~\ref{fig:sm-edge}(b) set the numerical background below which the
physical splitting is unresolved. They are compatible with the
number-conserving edge theory of a gapped neutral sector and a gapless
charge mode~\cite{RBA2015,KeselmanBerg2015}, but do not by themselves
determine an exponential localization length or splitting prefactor.

\section{Fixed-pairing comparison and scope}
\label{sm:scope}

\subsection{Bogoliubov--de Gennes reference}

A fixed-pairing Bogoliubov--de Gennes (BdG) Hamiltonian provides a useful
band-level reference but does not determine the transition of the
number-conserving interacting model. Add an on-site singlet amplitude
$\Delta$ to the rotated one-body Hamiltonian in
Eq.~\eqref{eq:sm-band}. At the particle--hole-invariant momenta
$k=0,\pi$, the odd-in-momentum term vanishes, and a gap closing occurs when
\begin{equation}
J^2=\xi_k^2+|\Delta|^2,\qquad
\xi_k=-2t\cos\frac Q2\cos k-\mu .
\label{eq:sm-bdgclosing}
\end{equation}
Here $\mu$ is the chemical potential of the fixed-pairing reference model.
The class-D interval is the region in which the two quantities
$J^2-\xi_0^2-|\Delta|^2$ and
$J^2-\xi_\pi^2-|\Delta|^2$ have opposite signs
~\cite{Kitaev2001,OregRefaelVonOppen2010,LutchynSauDasSarma2010}.
At $Q=\pi/2$ and density $n=1/2$, the $k=0$ band edge lies near the chemical
potential, reducing the lower fixed-$\Delta$ threshold.

For a fixed-density comparison, however, $\mu$ must be adjusted as $J$ and
$\Delta$ change. Moreover, the microscopic $U=3$ model has no imposed
pairing amplitude whose value can be inserted into
Eq.~\eqref{eq:sm-bdgclosing}. We therefore use the BdG calculation only to
state the band-closing criterion; together with
Eq.~\eqref{eq:sm-projectedpair}, it identifies the odd-parity intraband
pairing channel.
We do not identify its lower threshold with the interacting
$J_c\simeq0.77$, nor infer the direction of the many-body shift from the BdG
reference.

\subsection{Range of the exact and numerical statements}

The exact spectral relation in Eq.~\eqref{eq:sm-spectral} requires only a
static coplanar texture whose total winding is an integer and interactions
that are invariant under the local spin rotation. It is independent of
$U$, $J$, density, and the existence of topological pairing. The reduction
to a scalar boundary phase is special to the coplanar integer-winding case;
a generic noncoplanar texture has a matrix-valued $SU(2)$ boundary rotation.

The phase-dependent parity selection is a low-energy statement with
additional conditions. The ring must be long compared with the neutral
correlation length so that the difference between the two topological
neutral boundary sectors is exponentially small, and the winding energy
$\pi^2\rho_s/(2L)$ must be below the lowest neutral excitation in the
opposite-parity sector.
The numerical evidence presented here is for $n=1/2$, $U=3$, textures near
$Q=\pi/2$, and the topological comparison at $J=1.5$. The adjacent-winding
data at $J=1.5$ demonstrate the reversal and are consistent with the leading
pitch correction and its cancellation under symmetric averaging.
The trivial-phase comparison at $J=0.3$ applies the same adjacent-winding
protocol at $L=16$, $24$ and $32$, including the full flux interval at
$L=24$, and shows no reversal of either sign index
(Sec.~\ref{sm:winding:trivial}).

Finally, the texture is treated as a frozen classical background. Slow
texture dynamics, electronic feedback on the magnetic order, quasiparticle
poisoning, and finite-temperature spectroscopy are outside the Hamiltonian
studied here. They affect how the boundary-condition response would be
resolved experimentally, but not the fixed-texture unitary identity.

\end{document}